# Sensitive biodetection in flow using metasurface hosting quasi-bound state in the continuum resonances

*Sarah L. Walden*†‡§, Anna Fedotova‡§, Dmitry Pidgayko‡§, Katsuya Tanaka‡§‖, Chengjun Zou‡§, Thomas Pertsch§‖, Isabelle Staude‡§*

†School of Environment and Science, Griffith University, 170 Kessels Road, Nathan 4111, Australia. ‡Institute of Solid State Physics, Abbe Center of Photonics, Friedrich Schiller University Jena, Helmholtzweg 3, 07743, Jena, Germany. §Institute of Applied Physics, Abbe Center of Photonics, Friedrich Schiller University Jena, Albert-Einstein-Strasse 15, 07745, Jena, Germany. ‖Fraunhofer Institute for Applied Optics and Precision Engineering, Albert-Einstein-Str. 7, 07745 Jena, Germany

*Corresponding Email: s.walden@griffith.edu.au*

ABSTRACT: We have designed optical metasurfaces hosting high-quality factor quasi-bound state in the continuum (q-BIC) resonances for optical biosensing in flow. The unit cell of the metasurface contains two rectangular bars. An asymmetry factor is introduced by varying the gap width between the bars, to enable optical coupling to a q-BIC resonance confined to the air gap between neighboring nanoresonators. The location of the resonances makes them highly sensitive to changes in the local refractive index, leading to experimental bulk refractive index sensitivities exceeding 315 ± 22 nm/RIU and a figure-of-merit of 66 ± 5 RIU$^{-1}$. Successful streptavidin-biotin binding was observed by measuring the metasurface transmission in real-time by exposing the metasurface to various concentrations of analytes via a commercial microfluidic flow cell apparatus. The experimental limit of detection, defined as 3σ above noise, was found to be $1.8x10^{-8}$ M. This platform represents a compact optical approach for point-of-care diagnostics with fast read-out.

**Main Text**

Optical metasurfaces hosting high-quality factor quasi-Bound States in the Continuum (q-BIC) are renowned for their exceptional ability to confine light[1]. The hallmark feature of q-BICs is the minor asymmetries introduced into the metasurface unit cell to allow far-field coupling to the resonant structures. The broken symmetry leads to narrow spectral features with high quality factors ($Q = \lambda/\Delta\lambda$) exceeding $10^5$ at the resonance wavelength ($\lambda$).[2] These narrow resonances are highly sensitive to the refractive index of the medium surrounding the nanoresonator and therefore offer a highly sensitive photonic platform for detecting biomolecules at low concentrations[3–5].

The effectiveness of a photonic sensor is primary dictated by the sensitivity $S = \delta\lambda_0/\delta n$ where $\lambda_0$ is the wavelength of the electromagnetic resonance and $n$ is the refractive index of the sensing medium. Typical sensitivity values for dielectric metasurfaces range from S = 100-700 nm/RIU,[6,7] with plasmonic metasurfaces achieving values up to S = 1000 nm/RIU.[8] The significantly higher sensitivities achieved from plasmonic structures arises due to the localization of electric field on the surface of the particle (depicted in Fig 1), in the region where the target analyte is located. Plasmonic structures, however, suffer from low quality factors that reduce their practical detection limit. Mie resonances, in contrast to plasmonic resonances, have the advantage of negligible losses at the resonant wavelength, but have the critical drawback that the electric field enhancement is located within the high refractive index medium (depicted in Fig 1).[9] Our approach is to use employ a unit cell containing two rectangular nanoresonators (2-bar), with small asymmetries in the structure to introduce q-BIC resonances that are localized between neighboring nanoresonators. This design combines the high refractive index and low losses of dielectrics with the resonances located within the sensing volume pictured in Fig. 1.

Similar q-BIC metasurfaces have been demonstrated for optical sensing based on a variety of geometries including 2-bar[10–14], 3-bar[15], tilted ellipses[16–19], crescent structures[20], dual ribbons[21], slotted[22,23] or full disks[24,25]. Yet, with current trends toward wearable sensors and point-of-care diagnostics, the application of q-BIC metasurfaces for real-time biosensing[26,27] remains underexplored. Previous examples of q-BIC metasurfaces for detection of streptavidin proteins report sensitivities down to 5 ng/mL,[28] however the vast majority of these works perform incubation,[17,20,28] multiplexing[16,19,6], amplification steps[29,30] or machine learning algorithms[31] to enhance the detection limit. For practical applications, high sensitivities are required when measurements are performed in real-time without additional amplification. Recently, our group has reported 2-bar dielectric metasurfaces hosting several q-BIC resonances as a platform for metasurface tuning applications[32,33]. But any system whose behavior can be tuned by external stimuli, can also be employed for sensing applications.

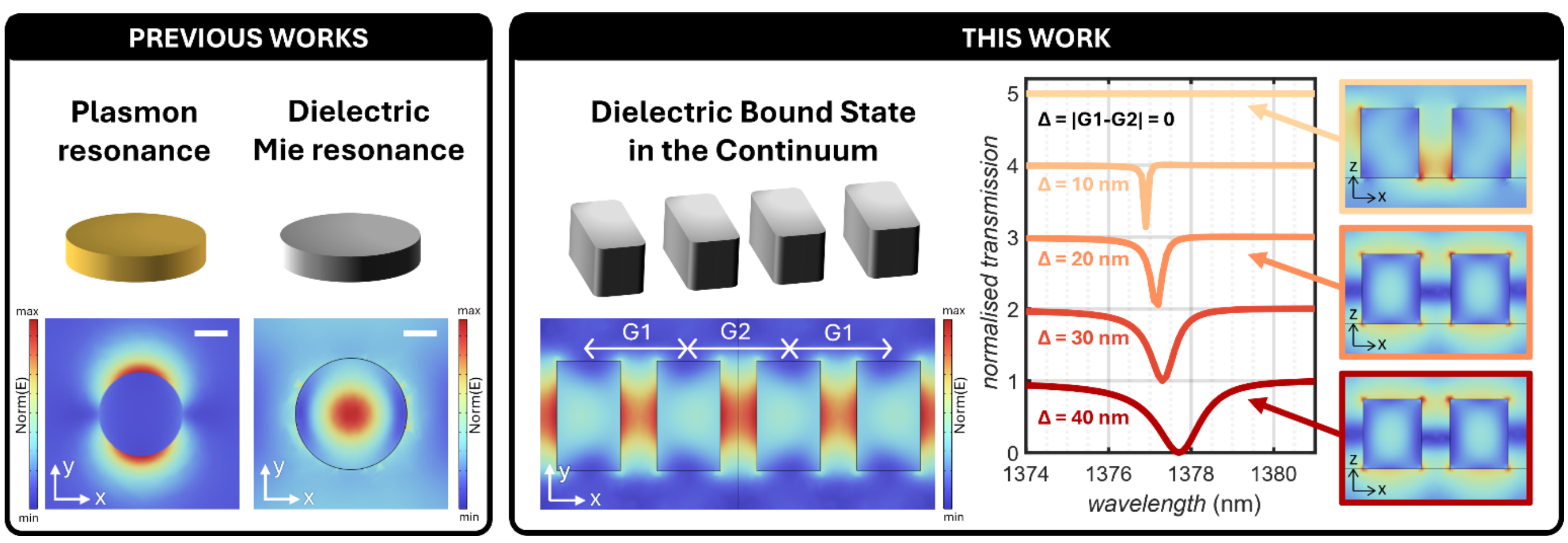

**Figure 1**. (left panel) Simulated electric field distributions demonstrating localization of plasmonic resonances to the surface of metallic nanostructures and Mie resonances within dielectric nanostructures.

Scale bar indicates 100 nm. (right panel) BIC resonances in 2-bar dielectric nanostructures enable mode localization within the sensing volume. Breaking the symmetry of the unit cell structure by varying the gap width between nanoresonators (G1) and between nanoresonators of neighboring unit cells (G2) ($\Delta = |G1 - G2|$) enables coupling to the optical mode leading to sharp dips in the transmission spectrum. Transmission spectra have been vertically offset by 1 for clarity.

The unit cell of the metasurface array used in this work has a period of Λ = 770 nm and comprises of two silicon nanoresonators, with equal width and height, separated by a center-to-center distance of $G1 = 360$ nm. Structural asymmetries are introduced into the unit cell, by varying the length of the bars and gaps between neighboring bars, enabling light to couple to the q-BIC resonances. The simulated transmission spectra across the entire telecom band, with a description of the resonant modes, are provided in Section 3.1 of the Supporting Information. The key resonance for this work is depicted in Figure 1 (left) and arises from the differing gap widths between the constituent nanoresonators within the unit cell ($G1$) and the nanoresonators from neighboring unit cells ($G2$). Simulations of the electromagnetic field distributions and transmission spectra for various values of $\Delta = |G1 - G2|$ are presented in Figure 1. When $\Delta = 0$ ($G1 = G2 = 385$ nm) the structure is symmetric and no light can couple into the BIC resonance, leading to near-unity transmission at the resonance wavelength. As $\Delta$ increases the width of the q-BIC resonances increases while maintaining strong electric field enhancement. A value of $G1 = 360$ nm was selected for the fabricated structures to ensure the resonance was detectable within instrument limitations.

The metasurface arrays were fabricated by an electron beam lithography (EBL) procedure described in detail in Section 1.1 of the Supporting Information. Briefly, amorphous silicon thin films were prepared and coated with a conductive chromium layer. EBL was then employed using a negative tone electron-beam resist to fabricate 600 μm x 600 μm metasurface arrays with an EBL dose of 110 μC/cm$^2$. After EBL, reactive ion etching removed silicon in the un-exposed areas leaving only the nanoresonators. Scanning electron microscopy (SEM) images of the fabricated structures are provided in Figure S1 of the Supporting Information.

The dimensions of the fabricated structures were determined by comparing finite element method simulations in the commercial software package Comsol Multiphysics with experimental transmission spectra and SEM images. Comparative spectra are provided in Figure S5 of the Supporting Information. The fabricated structures were determined to have equal widths W = 242 nm and heights H = 270 nm, but differing lengths ($L_1$ = 405 nm, $L_2$ = 373 nm). As stated above, the key resonance for this work is the in-plane electric dipole resonance located between the two nanoresonators of the unit cell (refer to Section 3.1 of the Supporting Information). The Q-factor of the resonance was experimentally determined to be Q = 141, but increased to Q = 261 when the metasurface was immersed into Phosphate-Buffered Saline (PBS) solution due to the reduced contrast between the refractive indices of the solution and substrate[34]. A similar resonance was observed between nanoresonators for y-polarized incident light, but this resonance overlaps with a broad magnetic dipole resonance, limiting its practical use for sensing. For full details refer to Section 3.1 of the Supporting Information.

To demonstrate the efficacy of our metasurface design for biosensing, we start by characterizing the bulk refractive index sensitivity. This is done by flowing known mixtures of ethanol:water over the metasurface mounted in a microfluidic cell and measuring the transmission in situ using a custom built apparatus outlined in detail in Section 1.3 of the Supporting Information. The measured metasurface transmission spectra for each solvent mixture are presented in Fig. 2A. A linear fit of the resonance shift for small changes in refractive index shown in Fig. 2B revealed a bulk refractive index sensitivity[35] of $S = \delta\lambda_0/\delta n = 315 \pm 22$ nm RIU$^{-1}$ and a figure-of-merit of $FOM = S/\Delta\lambda = 66 \pm 5$ RIU$^{-1}$. The experimental sensitivity is slightly lower than the theoretical value of 358 nm $RIU^{-1}$ determined from simulations, likely due to fabrication tolerances and complexities of fluid dynamics in nanoscale confined environments. To demonstrate the benefit of employing a flow setup for sensing, we further present kinetic measurements showing the real-time shift in the minima of the q-BIC resonance as the sample is exposed to solutions of different refractive index. For these measurements, transmission spectra were recorded every 18 s as the sample is exposed to different solvent mixtures. Starting with pure water, each solvent mixture was flowed across the metasurface at a rate of 50 µL/min for 10 minutes. The results, presented in Fig. 2C, clearly show a visible shift in the resonance for ethanol:water mixtures as low as 1% vol/vol. The shift is completely recovered when the sample is once again exposed to pure water. As higher concentration solutions are flushed across the surface the magnitude of the shift increases, but in all cases recovers back to the original position.

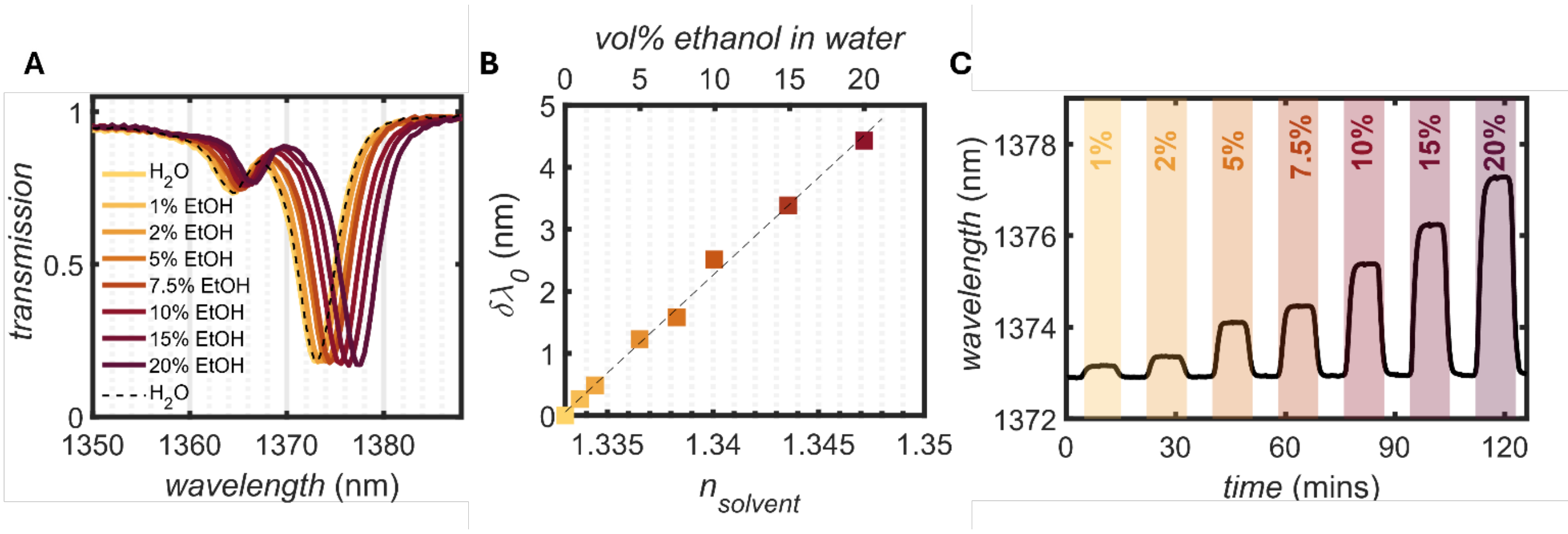

**Figure 2**. (A) Measured shift in metasurface transmission spectra when exposed to different ethanol:water solvent mixtures, before finally returning to pure water (dashed black line). (B) Linear fit of the resonance shift for small changes in refractive index. Gradient of fit was used to determine the bulk refractive index sensitivity. (C) Kinetic measurements showing the shift in the minima of the q-BIC resonance as the metasurface is exposed to various solutions of ethanol:water. Spectra were recorded

every 18 s as solutions passed through the flow cell at a rate of 50 μL/min. Shaded regions indicate the time period where the metasurface is exposed to the indicated solvent mixture. White regions are pure water.

With the bulk refractive index sensitivity established, the next step was to verify the biosensing capability of the metasurface. For these measurements, the metasurface was first functionalized with commercially available biotin-PEG-silane receptor and then exposed to solutions of streptavidin proteins dispersed in PBS buffer solution in a microfluidic cell as depicted in Scheme 1.

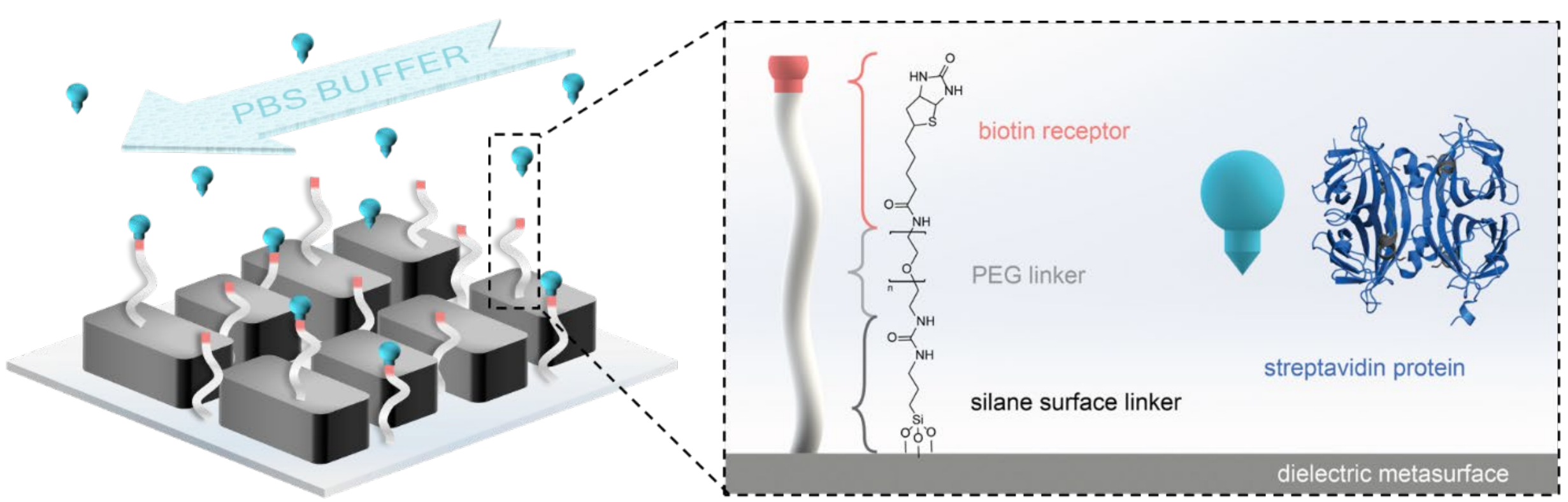

**Scheme 1.** Schematic diagram of metasurface functionalized with the biotin receptor binding to streptavidin proteins in PBS buffer solution resulting in detectable shift in q-BIC resonance. The zoomed-in region shows the chemical structure of the biotin-PEG-silane surface receptor and the streptavidin protein analyte. Streptavidin structure sourced from ref [36].

A full description of the surface functionalization procedure is provided in Section 1.4 of the Supporting Information. Briefly, 500 μL of biotin linker solution was drop cast onto the oxygen plasma activated metasurface and left to incubate for 1.5 hours before being washed off. Successful functionalization was confirmed by observing a small, but significant, shift in the q-BIC resonance. Biosensing was initially tested with solutions of streptavidin functionalized with quantum dots (strep-QD), in order to visually confirm successful binding and correlate the resonance shift with an increase in fluorescence intensity. Solutions of strep-QD were prepared in PBS buffer solutions at various concentrations ranging from $10^{-6}$ M to $10^{-3}$ M. The metasurface was mounted into the microfluidic cell and filled with PBS solution. For each concentration, 500 μL of the prepared solution was injected into the microfluidic cell at 50 uL/min, followed by PBS buffer solution to wash away any unreacted analyte. The metasurface transmission was measured at regular intervals during exposure and the results are presented in Fig. 3. The specificity of the binding to surfaces functionalized with biotin was confirmed and the results are included in Section 3.3 of the Supporting Information.

Fig. 3A shows the resonance shift after exposure to each concentration of strep-QD, with the inset highlighting the strong redshift in the q-BIC resonance with increasing strep-QD concentration. After each solution, the intensity of the fluorescence spectrum arising from the quantum dots was recorded (Fig. 3B) under exposure from a 365 nm LED and compared with the resonance shift (Fig. 3C) indicating good qualitative agreement. The inset in Fig. 3C shows a photograph of the sample under UV light after the experiments clearly showing the strep-QD binding is confined to within the bounds of the diamond-shaped microfluidic cell. An attractive feature of the streptavidin-biotin binding reaction is that the binding is reversible when held at elevated temperatures (>70 °C) for several minutes[37]. This means that after use, the sensor can be cleaned and re-used without the need to remove the sample from the microfluidic cell or repeated surface functionalization procedures.

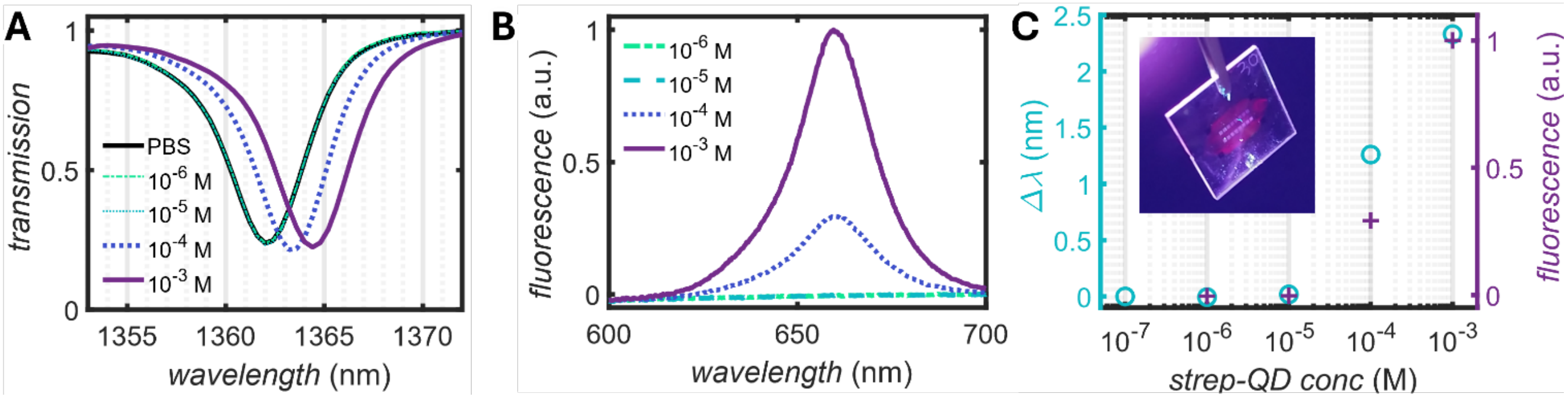

**Figure 3.** (A) Measured transmission spectra of the metasurface when exposed to different concentrations of strep-QD in PBS. Inset highlights the q-BIC resonance. (B) Fluorescence spectra of the strep-QD after each strep-QD solution and (C) comparison the fluorescence intensity and shift in the q-BIC resonance. Inset in (C) is a photograph of the metasurface under UV light after the experiment.

With the feasibility of the technique now well established, the sensitivity of the q-BIC metasurface for biodetection was measured. These experiments were identical to the abovementioned measurements on the strep-QD, but with pure streptavidin proteins from Streptomyces avidinii diluted to various concentrations between $10^{-8}$ M to $10^{-4}$ M in PBS solution. The metasurface was exposed to the samples in flow via a microfluidic chip at a flow rate of 50 μL $min^{-1}$ as the transmission was continually monitored. The results of the transmission spectra and kinetic measurements, tracking the wavelength minima of the transmission dip, are presented in Fig. 4. As expected, the q-BIC resonance undergoes a pronounced redshift as the streptavidin concentration was increased, which can be tracked in real-time using kinetic measurements. A shift in the resonance position can be observed for concentrations down to $3x10^{-8}$ M.

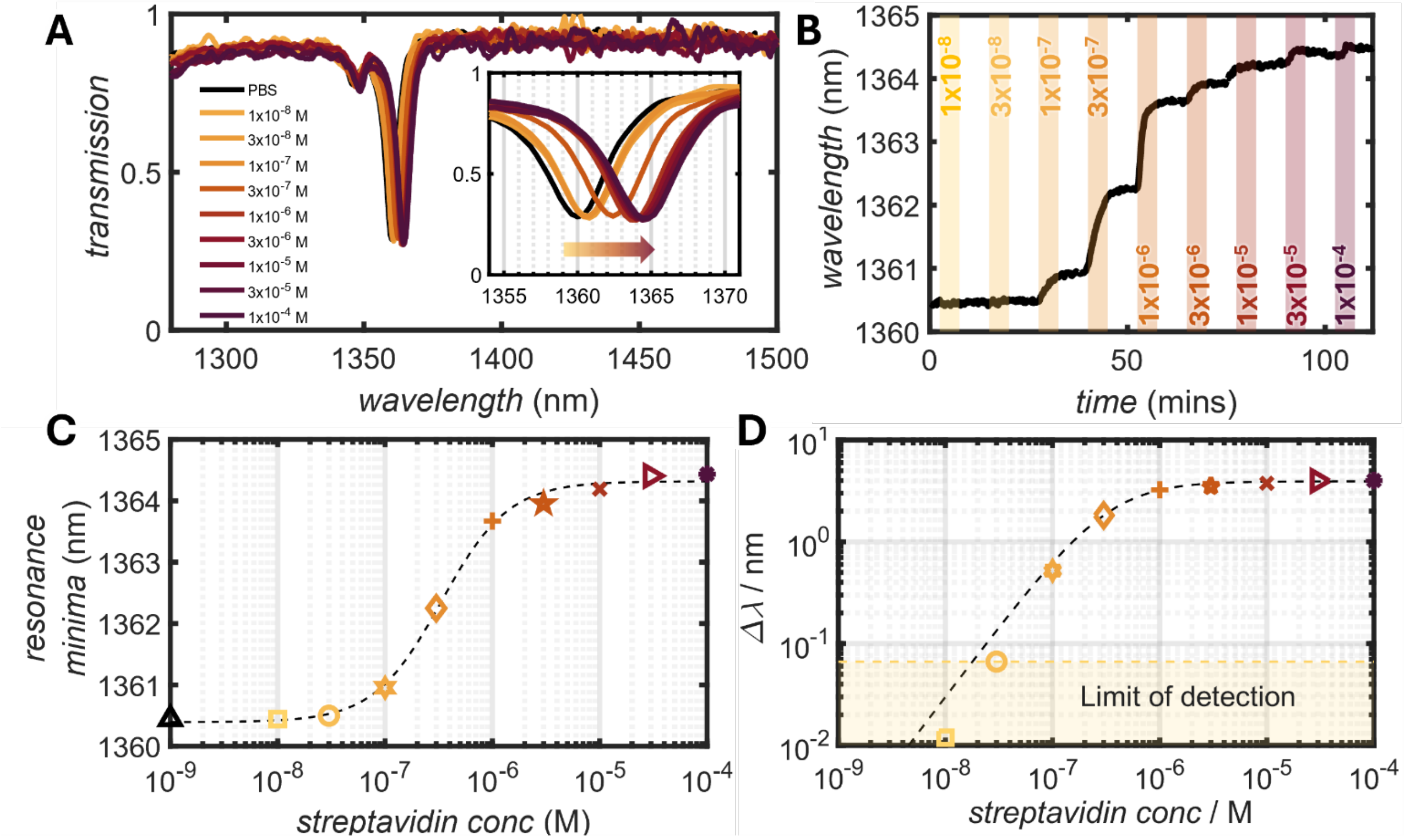

**Figure 4.** (A) Measured transmission spectra of the metasurface when exposed to different concentrations of strep-QD in PBS. The inset highlights the q-BIC resonance. (B) Kinetics of the shift in the q-BIC resonance over time as the metasurface is exposed to the streptavidin solutions in flow. (C) Hill plot of the sensing saturation at each concentration tested. Dashed line is fit to Hill Equation (Eqn. 1). (D) Same data as (C) plotted as resonance shift on a logarithmic scale with the 3σ limit of detection indicated.

The plateau resonance wavelength for each concentration was fit to the Hill equation[38,39] to assess the binding as a function of analyte concentration

$$\lambda = \lambda_0 + \frac{D}{1 + \left[10^{[C]-C_{1/2}}\right]^n} \tag{1}$$

where $\lambda$ is the wavelength minima at each concentration, $\lambda_0$ is the initial resonance position, $D = \lambda_\infty - \lambda_0$ is the dynamic range of the metasurface sensor, $[C]$ is the analyte concentration, $C_{1/2}$ is the concentration producing half occupation and $n$ is the Hill coefficient indicating the degree of cooperative or cascading of the reaction. Fitting Equation 1 to the experimental data produced good agreement as shown in Fig. 4C. The parameters $\lambda_0$, $D$, $C_{1/2}$ and n were found to be 1360.4±0.4 nm, 3.9±0.5 nm, $10^{-[6.5\pm0.2]}$ and 1.4±0.7, respectively. The limit of detection, determined as 3σ above the noise level[40] was $1.8\times10^{-8}$ M or 1.1 μg/mL. We expect that the limit of detection could be further improved with the addition of a reference channel[19,29], and a reduction in the channel depth to further localize the proteins in the vicinity of the metasurface.

In conclusion, we present here a platform for the sensitive detection of biomolecules in flow, based on dielectric metasurfaces hosting q-BIC resonances. The defining feature of this work is the localization of the q-BIC resonance in the gap between neighboring nanoresonators, making it highly sensitive to the local refractive index surrounding the nanoresonators. The bulk sensitivity of the metasurface arrays was experimentally determined to be 315 ± 22 nm $RIU^{-1}$. The metasurface was further demonstrated for biosensing in flow, showing detection of streptavidin molecules with a limit of detection of $1.8\times10^{-8}$ M. Our sensing platform represents a compact optical approach for point-of-care diagnostics with fast read-out in real-time.

AUTHOR INFORMATION

**Corresponding Author**

*Sarah L. Walden s.walden@griffith.edu.au, Isabelle Staude, isabelle.staude@uni-jena.de

**Author Contributions**

The manuscript was written through contributions of all authors. All authors have given approval to the final version of the manuscript.

**Acknowledgment**

This work was funded in part by the Deutsche Forschungsgemeinschaft (DFG, German Research Foundation) through the International Research Training Group (IRTG) 2675 "Meta-ACTIVE", project number 437527638. S.L.W. acknowledges funding from the Zukunfts Fellowship as part of the Jena Excellence Fellowship Program. The sample fabrication within this work was partly carried out by the microstructure technology team at IAP Jena. The authors thank them for providing the fabrication facilities, carrying out processes, and providing support.

**References**

(1) Koshelev, K.; Bogdanov, A.; Kivshar, Y. Meta-Optics and Bound States in the Continuum. *Science Bulletin* **2019**, *64* (12), 836–842. https://doi.org/10.1016/j.scib.2018.12.003.

(2) Hsu, C. W.; Zhen, B.; Stone, A. D.; Joannopoulos, J. D.; Soljačić, M. Bound States in the Continuum. *Nat Rev Mater* **2016**, *1* (9), 1–13. https://doi.org/10.1038/natrevmats.2016.48.

(3) Dell'Olio, F. Metasurface-Enabled Microphotonic Biosensors via BIC Modes. *Photonics* **2025**, *12* (1), 48. https://doi.org/10.3390/photonics12010048.

(4) Ghodrati, M.; Uniyal, A. Exploring Metasurface-Based Biosensor: New Frontiers in Sensitivity and Versatility for Biomedical Applications. *Plasmonics* **2024**. https://doi.org/10.1007/s11468-024-02640-7.

(5) Karawdeniya, B. I.; Damry, A. M.; Murugappan, K.; Manjunath, S.; Bandara, Y. M. N. D. Y.; Jackson, C. J.; Tricoli, A.; Neshev, D. Surface Functionalization and Texturing of Optical Metasurfaces for Sensing Applications. *Chem. Rev.* **2022**, *122* (19), 14990–15030. https://doi.org/10.1021/acs.chemrev.1c00990.

(6) Conteduca, D.; Barth, I.; Pitruzzello, G.; Reardon, C. P.; Martins, E. R.; Krauss, T. F. Dielectric Nanohole Array Metasurface for High-Resolution near-Field Sensing and Imaging. *Nat Commun* **2021**, *12* (1), 3293. https://doi.org/10.1038/s41467-021-23357-9.

(7) Wang, Y.; Ali, Md. A.; Chow, E. K. C.; Dong, L.; Lu, M. An Optofluidic Metasurface for Lateral Flow-through Detection of Breast Cancer Biomarker. *Biosensors and Bioelectronics* **2018**, *107*, 224–229. https://doi.org/10.1016/j.bios.2018.02.038.

(8) Kubo, W.; Fujikawa, S. Au Double Nanopillars with Nanogap for Plasmonic Sensor. *Nano Lett.* **2011**, *11* (1), 8–15. https://doi.org/10.1021/nl100787b.

(9) Chung, T.; Wang, H.; Cai, H. Dielectric Metasurfaces for Next-Generation Optical Biosensing: A Comparison with Plasmonic Sensing. *Nanotechnology* **2023**, *34* (40), 402001. https://doi.org/10.1088/1361-6528/ace117.

(10) Wang, T.; Liu, S.; Zhang, J.; Xu, L.; Yang, M.; Ma, D.; Jiang, S.; Jiao, Q.; Tan, X. Dual High-Q Fano Resonances Metasurfaces Excited by Asymmetric Dielectric Rods for Refractive Index Sensing. *Nanophotonics* **2024**, *13* (4), 463–475. https://doi.org/10.1515/nanoph-2023-0840.

(11) Dolia, V.; Balch, H. B.; Dagli, S.; Abdollahramezani, S.; Carr Delgado, H.; Moradifar, P.; Chang, K.; Stiber, A.; Safir, F.; Lawrence, M.; Hu, J.; Dionne, J. A. Very-Large-Scale-Integrated High Quality Factor Nanoantenna Pixels. *Nat. Nanotechnol.* **2024**, *19* (9), 1290–1298. https://doi.org/10.1038/s41565-024-01697-z.

(12) Hu, J.; Safir, F.; Chang, K.; Dagli, S.; Balch, H. B.; Abendroth, J. M.; Dixon, J.; Moradifar, P.; Dolia, V.; Sahoo, M. K.; Pinsky, B. A.; Jeffrey, S. S.; Lawrence, M.; Dionne, J. A. Rapid Genetic Screening with High Quality Factor Metasurfaces. *Nat Commun* **2023**, *14* (1), 4486. https://doi.org/10.1038/s41467-023-39721-w.

(13) Kühner, L.; Sortino, L.; Berté, R.; Wang, J.; Ren, H.; Maier, S. A.; Kivshar, Y.; Tittl, A. Radial Bound States in the Continuum for Polarization-Invariant Nanophotonics. *Nat Commun* **2022**, *13* (1), 4992. https://doi.org/10.1038/s41467-022-32697-z.

(14) Hakami, J.; Dhibi, A.; Bellouz, R.; Felidj, N.; Djaker, N. A High-Performance Biosensor Based on Quasi-Bound States in the Continuum for the Detection of Bacterial Species in Water. *Optics and Lasers in Engineering* **2026**, *196*, 109376. https://doi.org/10.1016/j.optlaseng.2025.109376.

(15) Huang, C.; Ma, Z.; Li, S.; Chen, X.; Zhang, H.; Lv, X.; Huang, B.; Chen, P.; Xie, Y. Self-Validating Quasi-BIC Metasensors for Reusable, Ultrasensitive, and Label-Free Biosensing. *Biosensors and Bioelectronics* **2026**, *296*, 118341. https://doi.org/10.1016/j.bios.2025.118341.

(16) Tittl, A.; Leitis, A.; Liu, M.; Yesilkoy, F.; Choi, D.-Y.; Neshev, D. N.; Kivshar, Y. S.; Altug, H. Imaging-Based Molecular Barcoding with Pixelated Dielectric Metasurfaces. *Science* **2018**, *360* (6393), 1105–1109. https://doi.org/10.1126/science.aas9768.

(17) Yesilkoy, F.; Arvelo, E. R.; Jahani, Y.; Liu, M.; Tittl, A.; Cevher, V.; Kivshar, Y.; Altug, H. Ultrasensitive Hyperspectral Imaging and Biodetection Enabled by Dielectric Metasurfaces. *Nat. Photonics* **2019**, *13* (6), 390–396. https://doi.org/10.1038/s41566-019-0394-6.

(18) Chen, W.; Li, M.; Zhang, W.; Chen, Y. Dual-Resonance Sensing for Environmental Refractive Index Based on Quasi-BIC States in All-Dielectric Metasurface. *Nanophotonics* **2023**, *12* (6), 1147–1157. https://doi.org/10.1515/nanoph-2022-0776.

(19) Jahani, Y.; Arvelo, E. R.; Yesilkoy, F.; Koshelev, K.; Cianciaruso, C.; De Palma, M.; Kivshar, Y.; Altug, H. Imaging-Based Spectrometer-Less Optofluidic Biosensors Based on Dielectric Metasurfaces for Detecting Extracellular Vesicles. *Nat Commun* **2021**, *12* (1), 3246. https://doi.org/10.1038/s41467-021-23257-y.

(20) Wang, J.; Kühne, J.; Karamanos, T.; Rockstuhl, C.; Maier, S. A.; Tittl, A. All-Dielectric Crescent Metasurface Sensor Driven by Bound States in the Continuum. *Advanced Functional Materials* **2021**, *31* (46), 2104652. https://doi.org/10.1002/adfm.202104652.

(21) Li, Y.; Lopez, E.; Plou, J.; Etxebarria-Elezgarai, J.; Pu, H.; Adam, J.; Chuvilin, A.; Seifert, A. High-Sensitivity Label-Free Biosensing Enabled by a Fabrication-Tolerant Dual-Ribbon Quasi-BIC Metasurface. *Advanced Optical Materials* **2026**, *14* (10), e03577. https://doi.org/10.1002/adom.202503577.

(22) Junhui W.; Deqiong L. I.; Guozheng N. I. E.; Jie Z.; Longfei G. a. N.; Zhiquan C.; Linfeng L. a. N. Near-infrared high-*Q* all-dielectric metasurface biosensor based on quasi-bound state in continuum. *Acta Phys. Sin.* **2025**, *74* (10), 107801–107809. https://doi.org/10.7498/aps.74.20241752.

(23) Zhang, X.; Wei, L.; Zhu, S.; Chen, Z. Ultrasensitive Metasurface Biosensor Based on Quasi-Bound States in the Continuum for Human Serum Albumin Detection. *Opt. Express, OE* **2025**, *33* (24), 50505–50518. https://doi.org/10.1364/OE.578467.

(24) Saadatmand, S. B.; Ahmadi, V.; Hamidi, S. M. Quasi-BIC Based All-Dielectric Metasurfaces for Ultra-Sensitive Refractive Index and Temperature Sensing. *Sci Rep* **2023**, *13* (1), 20625. https://doi.org/10.1038/s41598-023-48051-2.

(25) Loon, T. van; Liang, M.; Delplace, T.; Maes, B.; Murai, S.; Zijlstra, P.; Rivas, J. G. Refractive Index Sensing Using Quasi-Bound States in the Continuum in Silicon Metasurfaces. *Opt. Express, OE* **2024**, *32* (8), 14289–14299. https://doi.org/10.1364/OE.514787.

(26) Duarah, R.; Torné-Morató, H.; Zhang, G.; Amin, Y.; Pabast, M.; Sharma, N.; He, S.; Das, M. R.; Pompa, P. P. Next-Generation Wearable Optical Sensors for Personalized Health and Point-of-Care Diagnostics—A Systematic Review. *Advanced Healthcare Materials* **2026**, *15* (10), e04419. https://doi.org/10.1002/adhm.202504419.

(27) Wang, S.; Guan, X.; Sun, S. Microfluidic Biosensors: Enabling Advanced Disease Detection. *Sensors* **2025**, *25* (6), 1936. https://doi.org/10.3390/s25061936.

(28) Bontempi, N.; Chong, K. E.; Orton, H. W.; Staude, I.; Choi, D.-Y.; Alessandri, I.; Kivshar, Y. S.; Neshev, D. N. Highly Sensitive Biosensors Based on All-Dielectric Nanoresonators. *Nanoscale* **2017**, *9* (15), 4972–4980. https://doi.org/10.1039/C6NR07904K.

(29) Yavas, O.; Svedendahl, M.; Dobosz, P.; Sanz, V.; Quidant, R. On-a-Chip Biosensing Based on All-Dielectric Nanoresonators. *Nano Lett.* **2017**, *17* (7), 4421–4426. https://doi.org/10.1021/acs.nanolett.7b01518.

(30) Iwanaga, M. All-Dielectric Metasurface Fluorescence Biosensors for High-Sensitivity Antibody/Antigen Detection. *ACS Nano* **2020**, *14* (12), 17458–17467. https://doi.org/10.1021/acsnano.0c07722.

(31) Natraj, N. A.; Mubarakali, A.; Alagarsamy, M.; Al-Shamri, M. Y. H.; Dhivya, R. Next-Generation COVID-19 Detection Using a Metasurface Biosensor with Machine Learning-Enhanced Refractive Index Sensing. *Sci Rep* **2025**, *15* (1), 33406. https://doi.org/10.1038/s41598-025-18753-w.

(32) Zou, C.; Poudel, P.; Walden, S. L.; Tanaka, K.; Minovich, A.; Pertsch, T.; Schacher, F. H.; Staude, I. Multiresponsive Dielectric Metasurfaces Based on Dual Light- and Temperature-Responsive Copolymers. *Advanced Optical Materials* **2023**, *11*, 2202187. https://doi.org/10.1002/adom.202202187.

(33) Walden, S. L.; Poudel, P.; Zou, C.; Tanaka, K.; Paul, P.; Szeghalmi, A.; Siefke, T.; Pertsch, T.; Schacher, F. H.; Staude, I. Two-Color Spatially Resolved Tuning of Polymer-Coated Metasurfaces. *ACS Nano* **2024**, *18* (6), 5079–5088. https://doi.org/10.1021/acsnano.3c11760.

(34) Vincenti, M. A.; Carletti, L.; Ceglia, D. de; Rocco, D.; Weigand, H.; Saerens, G.; Falcone, V.; Grange, R.; Sedeh, H. B.; Li, W.; Litchinitser, N. M.; Scalora, M. From High- to Low-Contrast: The Role of Asymmetries in Dielectric Gratings Supporting Bound States in the Continuum. *Opt. Express, OE* **2024**, *32* (18), 31956–31964. https://doi.org/10.1364/OE.532622.

(35) White, I. M.; Fan, X. On the Performance Quantification of Resonant Refractive Index Sensors. *Opt. Express, OE* **2008**, *16* (2), 1020–1028. https://doi.org/10.1364/OE.16.001020.

(36) Le Trong, I.; Wang, Z.; Hyre, D. E.; Lybrand, T. P.; Stayton, P. S.; Stenkamp, R. E. Streptavidin and Its Biotin Complex at Atomic Resolution. *Acta Cryst D* **2011**, *67* (9), 813–821. https://doi.org/10.1107/S0907444911027806.

(37) Holmberg, A.; Blomstergren, A.; Nord, O.; Lukacs, M.; Lundeberg, J.; Uhlén, M. The Biotin-Streptavidin Interaction Can Be Reversibly Broken Using Water at Elevated Temperatures. *ELECTROPHORESIS* **2005**, *26* (3), 501–510. https://doi.org/10.1002/elps.200410070.

(38) Hill, A. V. The Possible Effects of the Aggregation of the Molecules of Haemoglobin on Its Dissociation Curves. *The Journal of Physiology* **1910**, *40*, 4–7. https://doi.org/10.1113/jphysiol.1910.sp001386.
(39) Gesztelyi, R.; Zsuga, J.; Kemeny-Beke, A.; Varga, B.; Juhasz, B.; Tosaki, A. The Hill Equation and the Origin of Quantitative Pharmacology. *Arch. Hist. Exact Sci.* **2012**, *66* (4), 427–438. https://doi.org/10.1007/s00407-012-0098-5.
(40) Conteduca, D.; Arruda, G. S.; Barth, I.; Wang, Y.; Krauss, T. F.; Martins, E. R. Beyond $Q$: The Importance of the Resonance Amplitude for Photonic Sensors. *ACS Photonics* **2022**, *9* (5), 1757–1763. https://doi.org/10.1021/acsphotonics.2c00188.

# Supporting Information

for

# Sensitive biodetection in flow using metasurface hosting quasi-bound state in the continuum resonances

*Sarah L. Walden*[†‡§]*, Anna Fedotova*[‡§]*, Dmitry Pidgayko*[‡§]*, Katsuya Tanaka*[‡§∥]*, Chengjun Zou*[‡§]*, Thomas Pertsch*[§∥]*, Isabelle Staude*[‡§]

[†]School of Environment and Science, Griffith University, 170 Kessels Road, Nathan 4111, Australia. [‡]Institute of Solid State Physics, Abbe Center of Photonics, Friedrich Schiller University Jena, Helmholtzweg 3, 07743, Jena, Germany. [§]Institute of Applied Physics, Abbe Center of Photonics, Friedrich Schiller University Jena, Albert-Einstein-Strasse 15, 07745, Jena, Germany. [∥]Fraunhofer Institute for Applied Optics and Precision Engineering, Albert-Einstein-Str. 7, 07745 Jena, Germany

*Corresponding Email:* s.walden@griffith.edu.au

## 1 Materials

- Biotin-PEG-Silane, MW 1k (#BP-24035) was purchased from Broadpharm,
- Cd-based Core/Shell Quantum Dots with Streptavidin (#OCNQSS665), $\lambda_{em}$ = 665 nm, 1 µM were purchased from Sigma Alrich GmbH
- Streptavidin from Streptomyces avidinii (#S4762-1MG) ≥13 units/mg protein was purchased from Sigma Alrich GmbH
- PBS (10X), pH 7.2 (#70013016) was purchased from Life Technologies GmbH
- All other reagents were used as received. All the solvents were of analytical grade.

## 1 Supplementary Experimental Methods

### *1.1 Metasurface Fabrication*

The silicon metasurfaces were fabricated using an electron beam lithography procedure in combination with reactive ion etching. Initially, commercially available amorphous silicon thin films situated on a glass substrate (Tafelmaier Dünnschicht Technik GmbH) were etched to the target silicon thickness of 270 nm to 280 nm by argon-ion beam etching (Oxford Instruments Plasma Technology, Ionfab 300), and covered with a conductive chromium layer by means of ion beam sputter deposition (Oxford Instruments Plasma Technology Ionfab 300). A negative electron-beam resist (100 nm EN038, Tokyo Ohka Kogyo Co., Ltd.) was then spin-coated onto the sample, and exposed by a variable-shaped electron-beam lithography system (Vistec SB 350OS) to define the areas to be selectively dissolved in a developer (OPD 4262) for 30 s. After transferring the exposed pattern to the chromium layer by ion-beam etching (Oxford Instruments Plasma Technology, Ionfab 300LC), the silicon layer was etched through the chromium mask by reactive ion etching (RIE-ICP, Sen-tech SI-500 C). Finally, the residual resist and chromium mask were effectively eliminated through a thorough washing process. Acetone was used to remove the resist, while a ceric ammonium nitrate-based solution, applied within a megasonic bath at 50 degrees Celsius, successfully eliminated the chromium mask.

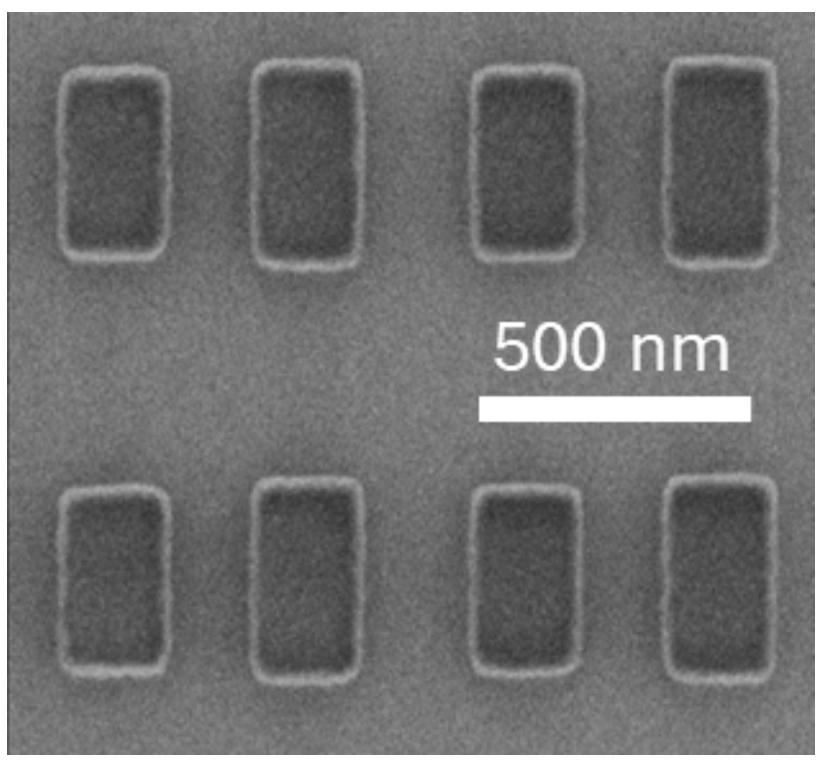

**Figure S1.** Scanning electron microscopy (SEM) image of fabricated metasurface arrays.

### *1.2 Metasurface Transmission Measurements*

The polarisation dependent transmittance spectra of the metasurfaces were measured using a custom-built optical setup depicted in Figure S2. The white light source was a tungsten halogen lamp (Thorlabs SLS301), fibre coupled and collimated with a condenser lens (f = 7 mm). The light was then transmitted through a polariser and incident on the sample. The transmitted light propagated through two apertures in the real and Fourier planes to limit the measurement area and incident angles, respectively. A removable mirror was used to select either the imaging path, where the light was directed to a CMOS camera (Thorlabs DCC1645C-HQ) for sample alignment. When the mirror is removed, the light is coupled into a fiber using a condenser lens (f = 20 mm) and directed to a Yokogawa AQ6370D optical spectrum analyzer to record the spectra at near infrared wavelengths.

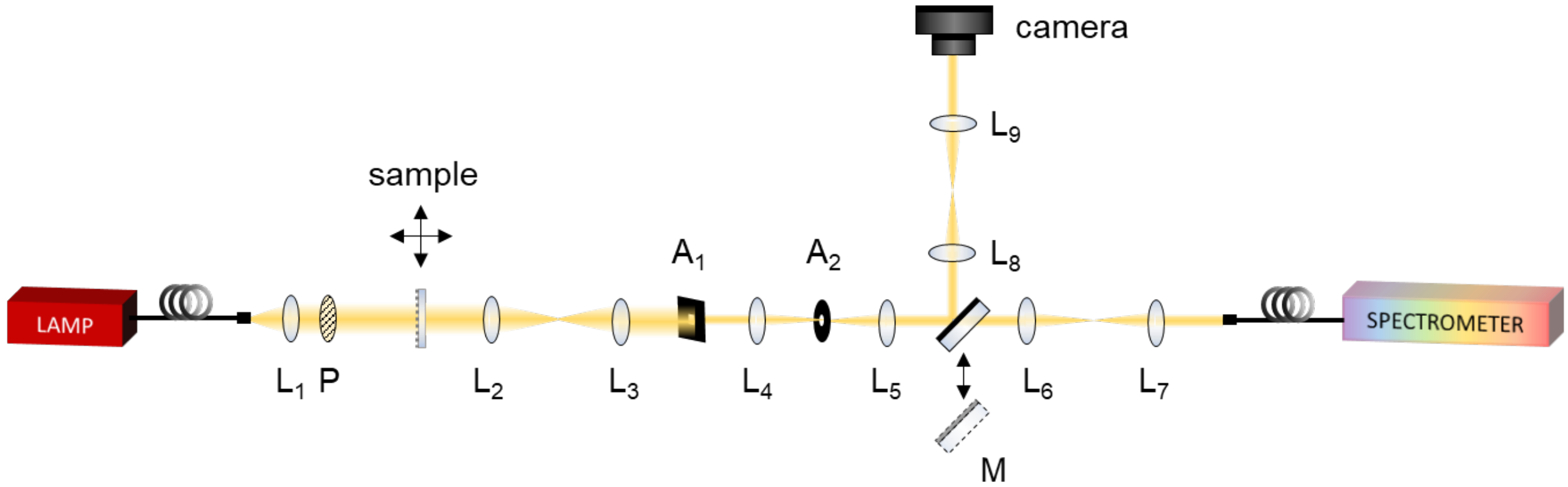

**Figure S2**. Optical setup for transmittance measurements under UV/blue light exposure. P indicates polarizer, M indicates mirror, $A_1$ and $A_2$ and rectangular and circular apertures, respectively. The focal lengths of the marked lenses are: $f_{L1}$ = 7 mm, $f_{L2}$ = 50 mm, $f_{L3}$ = 100 mm, $f_{L4}$ = 50 mm, $f_{L5}$ = 50 mm, $f_{L6}$ = 50 mm, and $f_{L7}$ = 200 mm.

*1.3 Bulk Refractive Index Sensing*

Bulk refractive index sensing was performed in flow using the transmission setup outlined in Section 1.2 and pictured in Figure S4. Prior to installation, the metasurface sample was cleaned with $O_2$ plasma for 5 mins at 100 W. The cleaned sample was then mounted into a Fluidic 864 sensor platform (#10001352) from Microfluidic ChipShop GmbH with the aid of an adhesive gasket tape. Various ethanol:water mixtures were flushed through the cell at a flow rate of 50 µL/min for 10 minutes, ensuring the cell volume had completely exchanged. Transmission measurements were measured every 18 s during flow. After 10 mins the sensing solution was replaced with water for 10 mins and then repeated with sensing solutions of various concentrations.

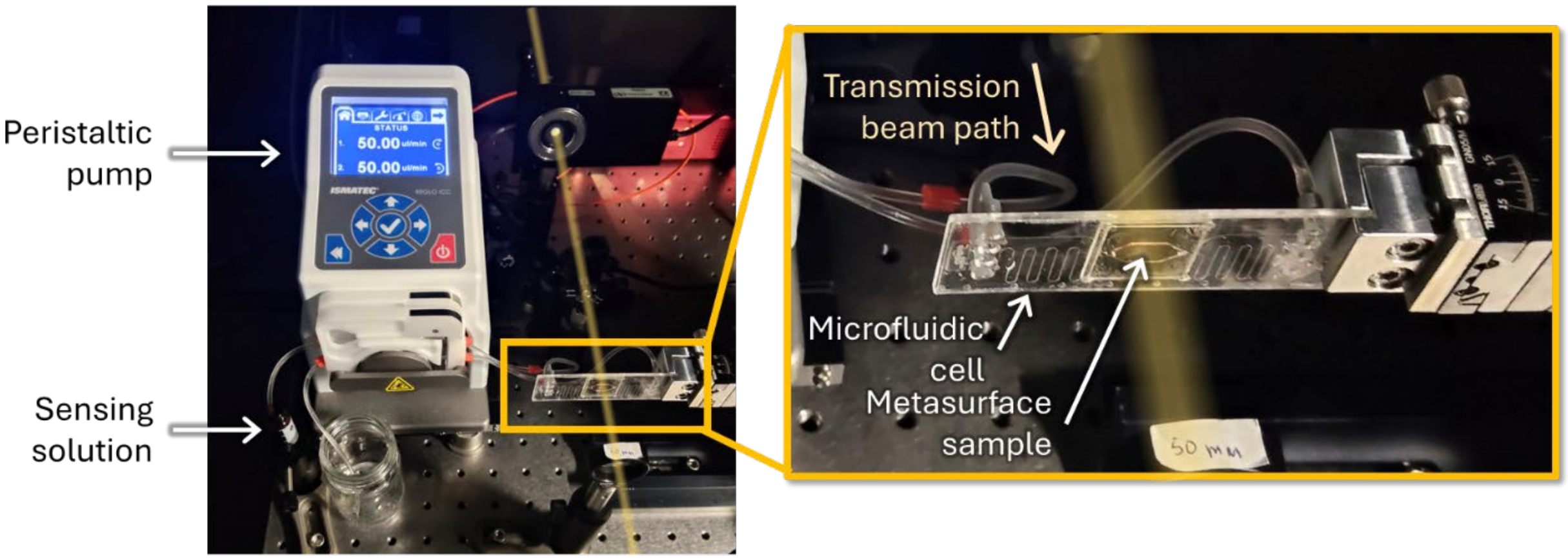

**Figure S3**. Photographs showing the experimental configuration for bulk refractive index sensing. (Main) Picture of the flow cell connected to the sensing solution via peristaltic pump. (Inset)

zoomed in picture of the metasurface sample mounted into the flow cell. The actual metasurfaces are not visible in the picture, as their footprint is only 600 µm x 600 µm.

*1.4 Biotin Surface Functionalization*

The metasurface was first exposed to oxygen plasma for 5 min at 150 W to activate the surface. 500 µL of a solution of the biotin linker in an ethanol:water mixture was drop cast onto the metasurface and left to incubate a room temperature for 1.5 hours. Afterwards the metasurface was rinsed sequentially in ethanol, water and isopropanol. Successful functionalization was confirmed by observing a small, but significant, shift in the q-BIC resonance.

*1.5 In-flow Biosensing Experiments*

Due to the smaller sample volumes, the flow cell used for these experiments was a Fluidic 864 sensor platform (#10001352) from Microfluidic ChipShop GmbH. The metasurface arrays were functionalized with biotin receptors using the procedure outlined in Section 1.4 and then the functionalized metasurface was installed into the microfluidic cell using 140 µm thick adhesive tape gaskets.

For measurement, the microfluidic cell with the metasurface installed was flushed with water and then pure PBS (1X) solution at a rate of 50 µL/min and the transmission spectrum of the metasurface was measured. The transmission spectrum was then continually measured every 15 s during flow to track any shift in the quasi-BIC resonances. While measurements were running, the various analyte solutions were introduced into the flow cell followed by pure PBS buffer solution until sensing saturated.

Initially quantum dots with streptavidin (strep-Qdots) were used for sensing. The purchased 1 µM stock solution was diluted with PBS buffer solution (1X) to prepare 200 µL solutions at $10^{-8}$ M, $5x10^{-8}$ M, $10^{-7}$ M, $5x10^{-7}$ M, $10^{-6}$ M $5x10^{-6}$ M, $10^{-5}$ M $5x10^{-5}$ M and $10^{-4}$ M. For pure streptavidin from Streptomyces avidinii (M = $6x10^{4}$ g/mol), the 1 mg powder was dispersed in 4.2 mL of PBS buffer solution (1X) to produce a $4x10^{-6}$ M stock solution. This was then diluted with PBS buffer solution (1X) to produce 200 µL solutions at the above-mentioned concentrations.

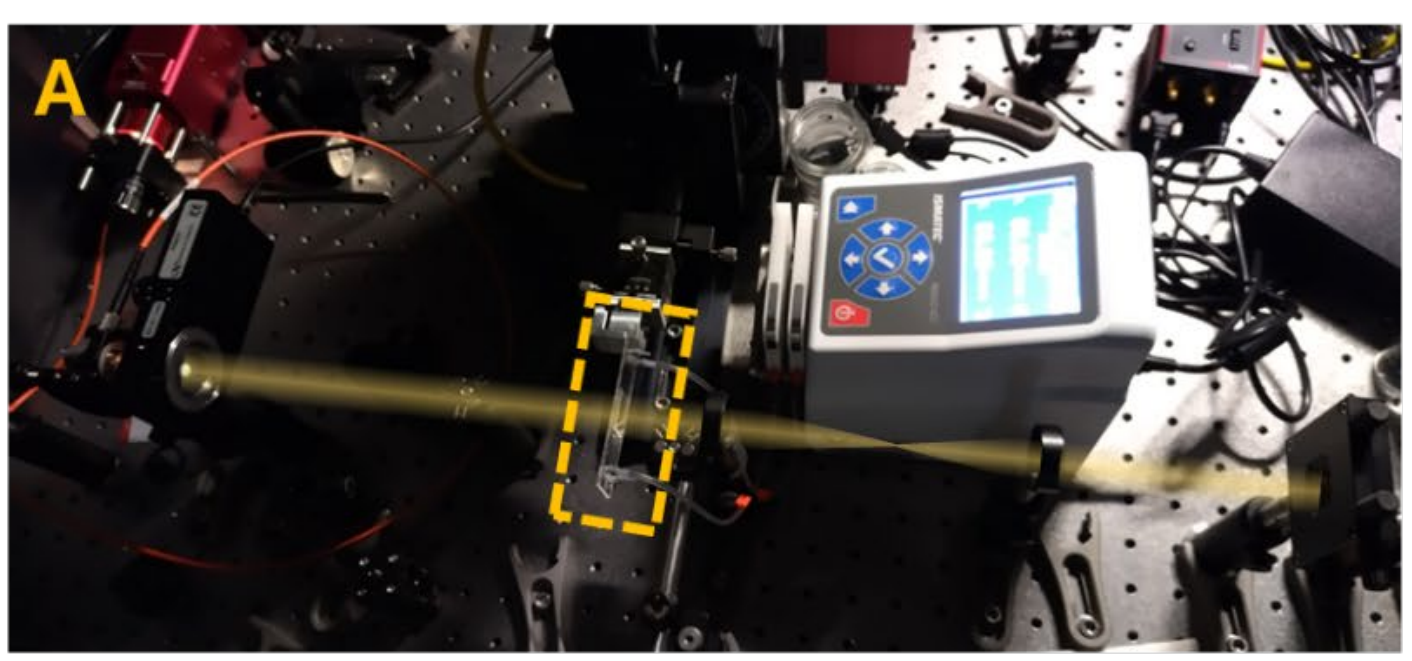

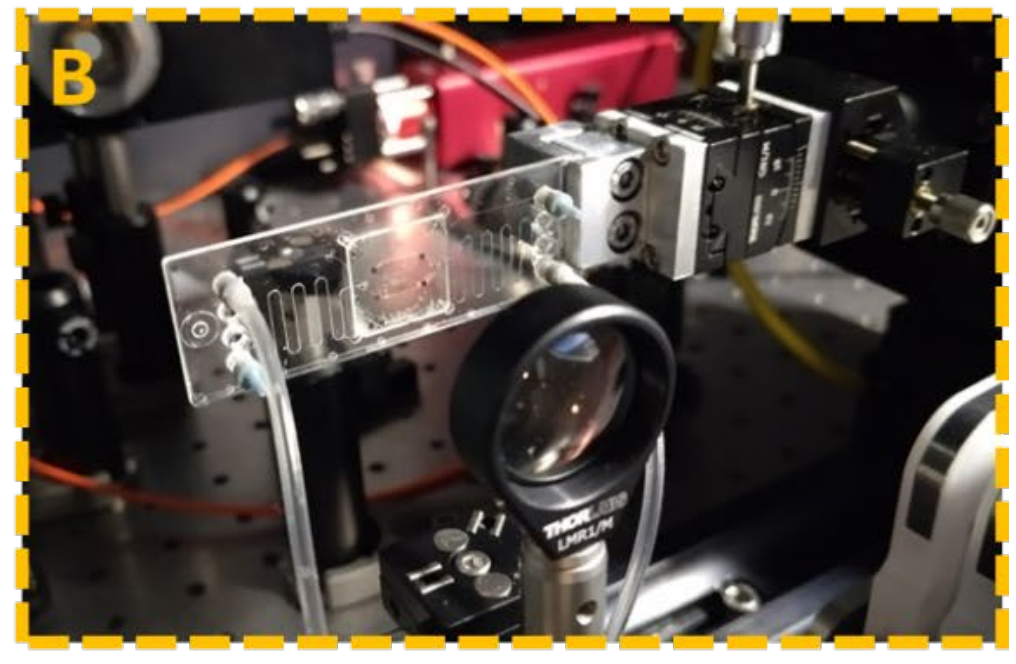

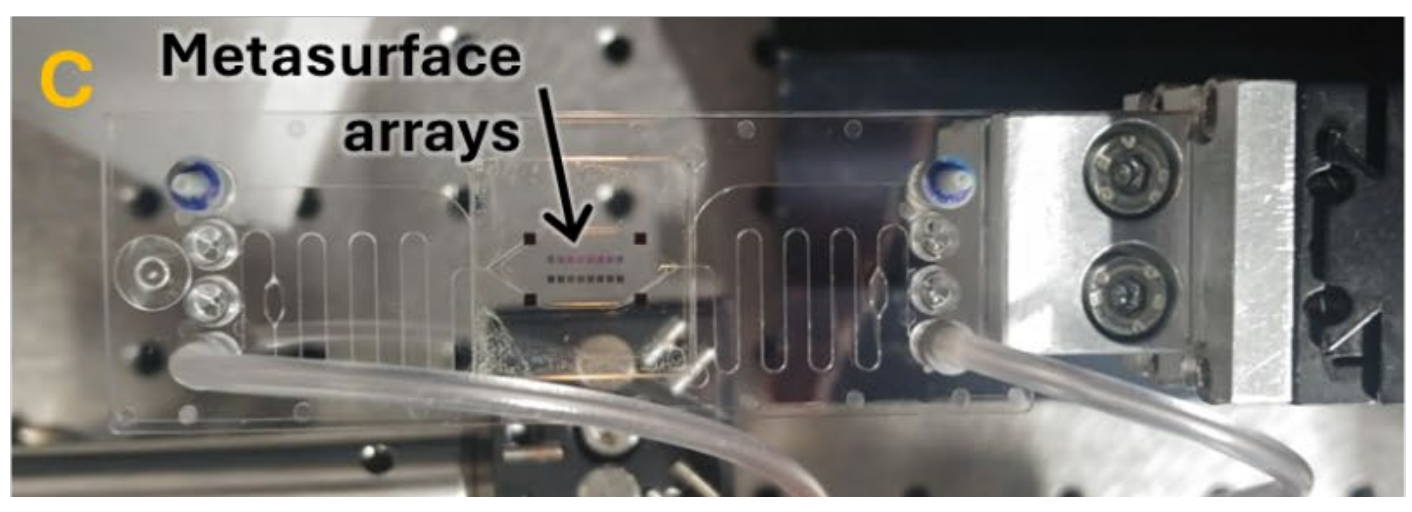

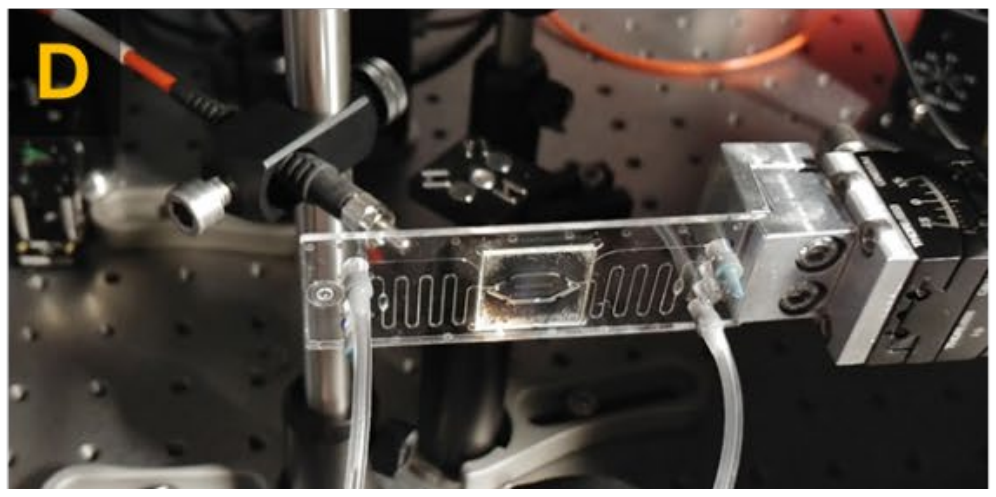

**Figure S4**. Photographs of various aspects of the biosensing apparats. (A) The sample mounted in position during kinetic transmission measurements. The rectangle indicates the metasurface sample position. The light path of transmission measurements is indicated by the transparent yellow line for clarity. (B) Zoomed-in picture of the flow cell position during measurement. (C) Close up picture of the sample containing the metasurface arrays mounted in the microfluidic cell. (D) position of optic fiber used to collect the fluoresced light from the strep-QD for fluorescence measurements.

## 3 Supplementary Data

### *3.1 Metasurface Mode Analysis*

Figure S5 presents the simulated transmittance spectra of the metasurface overlayed with the experimental spectra for x- and y-polarised incident light. Qualitatively, the experimentally measured and simulated spectra are in good agreement, despite fabrication tolerances meaning some of the q-BIC resonances are not as pronounced in the experimental data. To investigate the cause of the different resonances, the near-field electric and magnetic field profiles at the corresponding resonance wavelengths are presented in Fig. S5 (b-i). For the *x*-polarization, the relevant resonances are as follows:

- **X1:** the broad resonance observed around 1110 nm is due to antisymmetric, in-plane electric dipoles formed in each of the 2 bar resonators. The simulated electric field profile of resonance X1 is depicted in Fig. S5c.
- **X2:** The sharp resonance at 1180 nm is the q-BIC resonance formed by parallel, out-of plane electric dipole resonances formed in each of the 2 bar resonators. The simulated electric field profile of resonance X2 is depicted in Fig. S5d.
- **X3:** The dominant resonance located at 1240 nm is an antisymmetric, out-of-plane electric dipole pair, which induces a pronounced magnetic dipole moment between the two bars. The simulated electric field profile of resonance X3 is depicted in Fig. S5e.
- **X4:** The q-BIC resonance at 1320 nm arises due to antisymmetric, in-plane magnetic dipoles formed along the long axis of each of the 2 bar resonators. The simulated magnetic field profile of resonance X4 is depicted in Fig. S5f.

The metasurface transmission spectrum measured with y-polarised incident light is dominated by a broad magnetic dipole resonance centred around 1390 nm. Overlayed with this broad transmission dip are four q-BIC resonances with are outlined below.

- **Y1:** The resonance located at 1260 nm is due to parallel, out-of-plane magnetic dipole resonances formed in each of the 2 bar resonators, but is dominated by the gap between nanoresonators of neighbouring unit cells. This strong dependence on the gap distance, makes this resonance highly sensitive to the refractive index of this gap. The simulated magnetic field profile of resonance Y1 is depicted in Fig. S5g.
- **Y2:** The resonance located at 1290 nm corresponds to antiparallel, in-phase magnetic dipole resonances formed along the short axis of the bars. The simulated magnetic field profile of resonance Y2 is depicted in Fig. S5h.
- **Y3:** The resonance Y3, located at 1320 nm, is analogous to Y1, but dominated by the gap between the 2 bars of within the unit cell. The strong dependence on this gap makes this resonance highly sensitive to the surrounding environment. The simulated magnetic field profile of resonance Y3 is depicted in Fig. S5i.

- **Y4:** The resonance located at 1355 nm is an antisymmetric, out-of-plane magnetic dipole pair, which induces an electric dipole moment between the two bars. The simulated magnetic field profile of resonance Y4 is depicted in Fig. S5j.

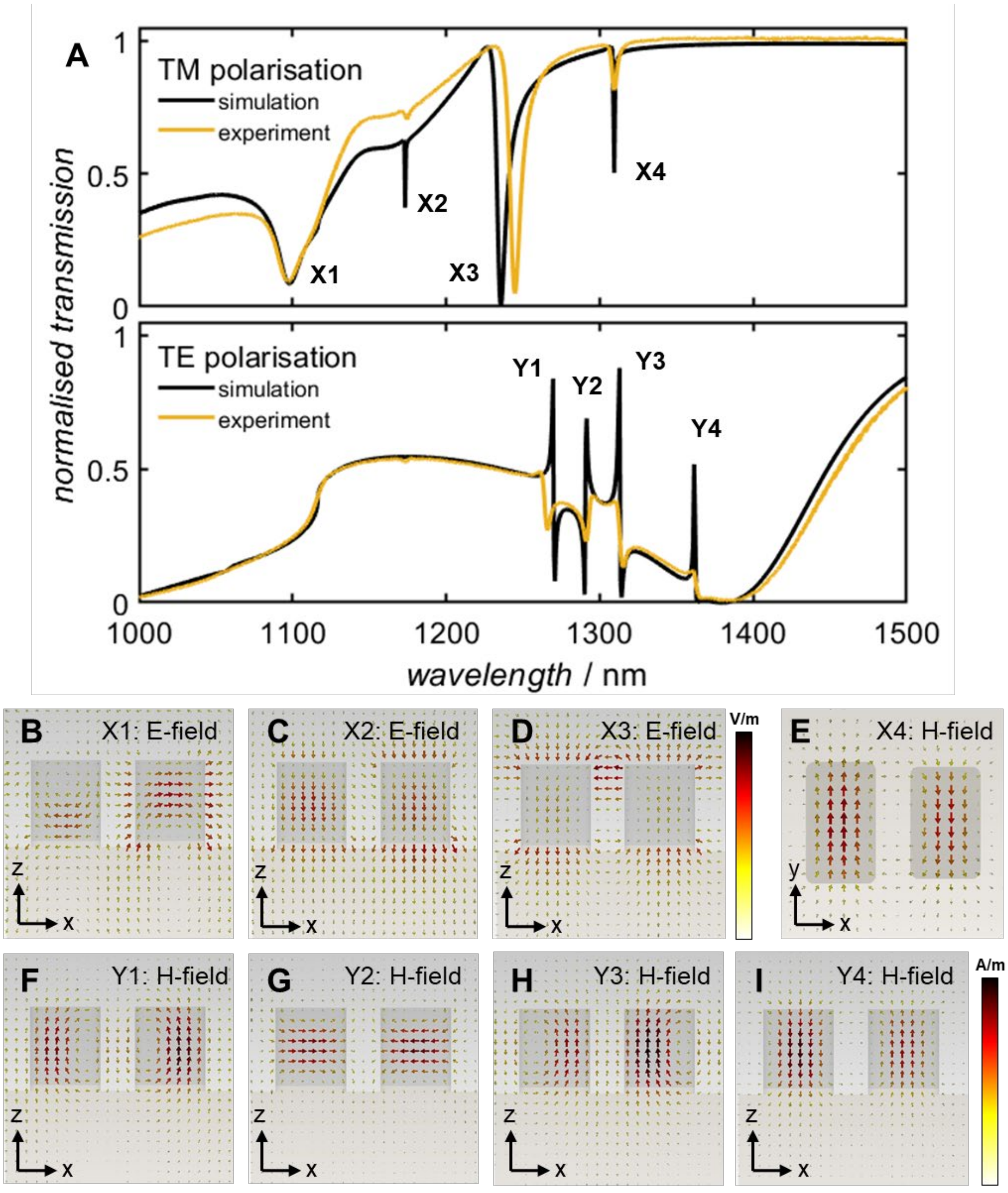

**Figure S5**. (A) Numerically calculated transmission spectra for (top) x- and (bottom) y-polarised incident light. (B-I) Numerically simulated near-field profiles of the unit cell for the sharp resonance features marked in (B).

*3.2 Simulations of Polymer-Coated Metasurface Resonance Tuning*

For all simulations, the two bar metasurface depicted in Figure 1 of the manuscript is employed. The relevant dimensions are period of unit cell $\Lambda$ = 770 nm, centre-to-centre gap between resonators G = 360 nm, length of long bar $L_1$ = 405 nm, length of short bar $L_2$ = 373 nm, width of both bars W = 242 nm and height of both bars $H_{bar}$ = 270 nm. A full discussion of the influence of these parameters on the resonance positions can be found in our earlier work.[3]

The calculated transmission spectra of the metasurface when coated with 500 nm thick layer with refractive index varying from 1.0 to 1.7 are presented in Figure S6. The top row presents data for x-polarised incident light and the bottom row is for y-polarised incident light. In both cases, the shifts of the four main resonances are indicated with a dashed grey line. For x-polarised light, the most significant shift is observed for X3, the anti-parallel electric dipole depicted in the near field profiles of Figure S6b and S6c. This transmission dip associated with resonance X3 shifts from $\lambda$ = 1235 nm when n = 1.0 to $\lambda$ = 1483 nm when n = 1.7, (average sensitivity of 354 nm $RIU^{-1}$) highlighting the sensitivity of this resonance to the local environment. The resonances X1 and X4, arise from the in-plane anti-parallel electric and magnetic dipoles, respectively. Both these resonances are highly localised within the nanobars, and hence only experience a weak shift in response to the refractive index change. Similarly, the transmission dip attributed to X2 arises from the parallel-out of plane electric dipoles located within the nanobars, which are largely insensitive to the refractive index change of the layer.

The bottom row of Figure S6, shows the transmission of y-polarised light through the coated metasurface. The main resonance of interest to this work is the parallel out-of-plane magnetic dipole resonance Y3 located between the two bars of the unit cell. The transmission dip attributed to this resonance undergoes a significant shift from $\lambda$ = 1313 nm when n = 1.0 to $\lambda$ = 1644 nm when n = 1.7, corresponding to an average sensitivity of 473 nm $RIU^{-1}$, significantly larger than that of resonance X3. The Q factor of this resonance also increases as the refractive index increases, and the resonance becomes localised between the nanobars. Resonance Y1 is physically the equivalent to resonance Y3, but is located between the nanobars of neighbouring unit cells, as opposed to between the two nanobars within a single unit cell. As a result, this resonance also undergoes a significant shift with refractive index change. Resonance Y2 and Y4 are anti-parallel in-plane and out-of-plane magnetic dipole resonances, respectively. These resonances are both highly localised within the nanobars and hence show minimal shift with layer refractive index change.

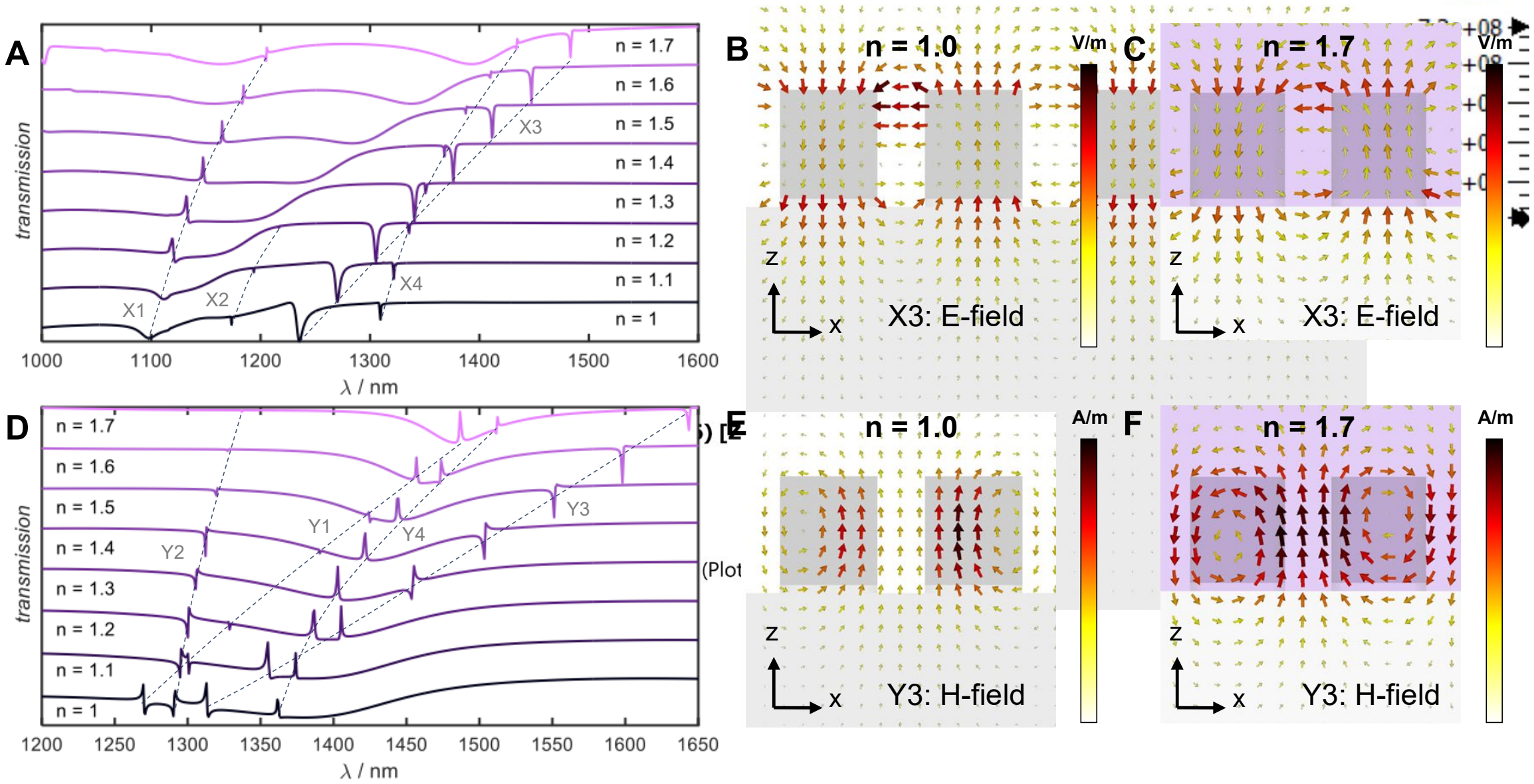

**Figure S6**. Metasurface simulations for a variation of the dielectric environment. (A,D) simulations showing expected electromagnetic resonances when the two bar metasurface is coated with a 500 nm thick layer with varying refractive index and illuminated with (A) x- and (D) y-polarised light. Grey dashed lines are a guide to the eye only, highlighting the shift of each resonance. (B,E) Electric near-field profiles and (E,F) magnetic near-field profiles in the unit cell of the metasurface when coated with a layer of refractive index (B,E) n=1 and (C,F) n=1.7.

*3.3 Specificity of Biotin Binding*

The specificity of the biotin-streptividin binding was confirmed via the drop-cast method. 100 uL of strep-QDs in PBS buffer solutions at various concentrations were drop-cast onto the bare metasurface (which have not undergone biotin-functionalization) and left to incubate for 1 hr. The sample was then rinsed with water and ethanol and the transmission of the metasurface was recorded. This was repeated for several different concentrations of strep-QD ranging from $10^{-7}$ M to $10^{-4}$ M. Finally, the entire process was repeated after the metasurface had been $O_2$-plasma cleaned and functionalized with biotin. The results, presented in Figure S7, confirm that a significant shift in the q-BIC resonance is only observed when the sample is functionalized with the biotin linker.

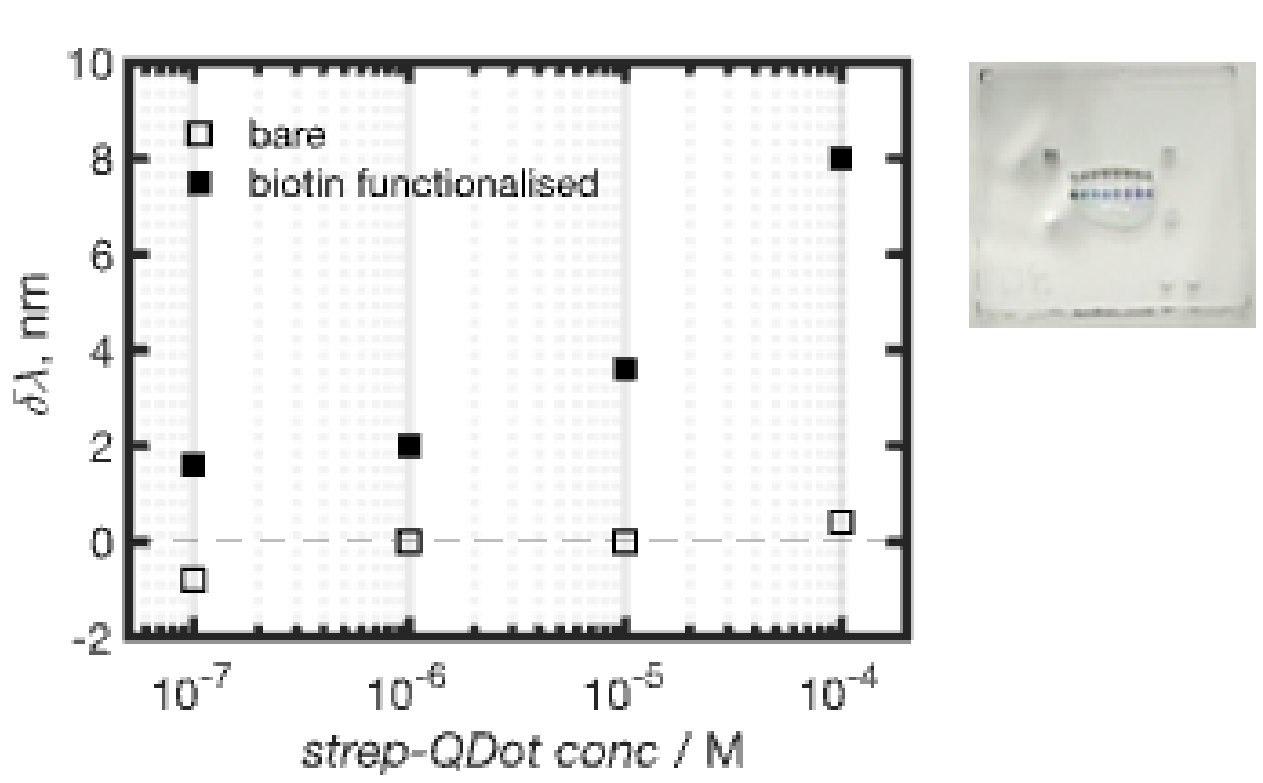

**Figure S7.** Measured shift in the q-BIC resonance with (solid squares) and without (hollow squares) functionalization with biotin linker. Photographs on right show the metasurface arrays incubated in strep-QD solutions for 1 hr and then washed before measurement.

References

(1) Shimoboji, T.; Ding, Z. L.; Stayton, P. S.; Hoffman, A. S. Photoswitching of Ligand Association with a Photoresponsive Polymer−Protein Conjugate. *Bioconjugate Chem.* **2002**, *13* (5), 915–919. https://doi.org/10.1021/bc010057q.

(2) Grimm, O.; Schacher, F. H. Dual Stimuli-Responsive P(NIPAAm-Co-SPA) Copolymers: Synthesis and Response in Solution and in Films. *Polymers* **2018**, *10* (6), 645. https://doi.org/10.3390/polym10060645.

(3) Zou, C.; Poudel, P.; Walden, S. L.; Tanaka, K.; Minovich, A.; Pertsch, T.; Schacher, F. H.; Staude, I. Multiresponsive Dielectric Metasurfaces Based on Dual Light- and Temperature-Responsive Copolymers. *Advanced Optical Materials* **2023**, *11*, 2202187. https://doi.org/10.1002/adom.202202187.